\documentclass[12pt]{article}
\usepackage{amsmath}
\usepackage{mathtools}
\usepackage{amssymb}
\usepackage{physics}
\usepackage{ytableau}
\usepackage{braket}
\pdfoutput=1
\usepackage[pdftex]{graphicx}
\usepackage{float}
\numberwithin{equation}{section}
\usepackage{cite}

\begin{document}
\setlength{\baselineskip}{0.7cm}
\begin{titlepage}
\begin{flushright}
NITEP 193 \\
KYUSHU-HET-275
\end{flushright}
\vspace*{10mm}%
\begin{center}{\Large\bf
Electroweak Symmetry Breaking \\
\vspace*{2mm}
in Two Higgs Doublet Model \\ 
\vspace*{2mm}
from 6D Gauge-Higgs Unification on $T^2/Z_2$ \\
}
\end{center}
\vspace*{10mm}
\begin{center}
{\large Kento Akamatsu}$^{a}$,
{\large Takuya Hirose}$^{b}$, 
{\large Nobuhito Maru}$^{a,c}$ and
{\large Akio Nago}$^{a}$ 
\end{center}
\vspace*{0.2cm}
\begin{center}
${}^{a}${\it
Department of Physics, Osaka Metropolitan University, \\
Osaka 558-8585, Japan}
\\
${}^{b}${\it Department of Physics, Kyusyu University, \\
Fukuoka 819-0395, Japan}
\\
${}^{c}${\it Nambu Yoichiro Institute of Theoretical and Experimental Physics (NITEP), \\
Osaka Metropolitan University,
Osaka 558-8585, Japan}
\end{center}
\vspace*{1cm}

\begin{abstract}
\end{abstract}
Electroweak symmetry breaking is explored in a two Higgs doublet model 
based on a six dimensional $SU(4)$ gauge-Higgs unification compactified on an orbifold $T^2/Z_2$. 
The remarkable property of this model is a prediction of realistic weak mixing angle $\sin^2 \theta_W = 1/4$ 
at the compactification scale. 
We calculate one-loop effective potential of the Standard Model Higgs boson 
from the contributions of the gauge boson 
and the fermion in a four-rank totally symmetric tensor where top quark is included.  
We find that the electroweak symmetry breaking certainly takes place.  
\end{titlepage}

\newpage
\section{Introduction}
Two Higgs Doublet Model (2HDM) is a simple extension of the Standard Model (SM), 
where an additional Higgs doublet is introduced to the SM \cite{2HDMReview} and 
its experimental aspects are also extensively studied \cite{2HDMexp}. 
Since the 2HDM predicts five physical Higgs scalars and a related phenomenology is very rich, 
much attention has been paid so far.  
Although the 2HDM is very attractive, some assumptions in tree level potential 
are usually proposed for phenomenological problems. 
First, a softly broken $Z_2$ symmetry is assumed to avoid 
Flavor Changing Neutral Current (FCNC) processes at tree level. 
Second, CP symmetry is assumed. 
Third, a degeneracy of mass is assumed to satisfy the constraint for the $\rho$ parameter. 
Furthermore, the gauge hierarchy problem is not solved and the Higgs mass cannot be predicted in 2HDM. 

Gauge-Higgs unification (GHU) is one of the interesting scenarios solving the gauge hierarchy problem, 
where the SM Higgs field is originated from the extra spatial component of the higher dimensional gauge field 
\cite{Manton, Fairlie, Hosotani1, Hosotani2}. 
In this scenario, the ultraviolet sensitivity of the Higgs sector is softened due to the higher dimensional gauge symmetry. 
One-loop effective potential of Higgs field is finite regardless of the non-renormalizable theory, 
 since it is described by non-local Wilson-line operators. 
Therefore, once the theory is fixed, the Higgs potential is calculable and Higgs mass can be predicted 
\cite{HIL, ABQ, MY, HMTY, LMH}. 

In a previous work by one of the authors \cite{HLM}, 
the 2HDM based on GHU was discussed. 
In order to realize 2HDM in GHU scenario, 
 six dimensional (6D) models compactified on an orbifold $T^2/Z_2$ are simplest \cite{SSSW}. 
In \cite{HLM}, the possibility of GHU models compatible with the realistic weak mixing angle 
$\sin^2 \theta_W = 1/4$ was explored. 
This is possible if the SM Higgs doublet field is embedded into $SU(3)$ triplet or sextet representation 
not an adjoint representation which is often considered. 
$G_2$ GHU model \cite{CGM, MS} and $SU(4)$ model \cite{MS, HLM} are known in the triplet case, 
on the other hand, $Sp(6)$ GHU model \cite{MS, HLM} is known in the sextet case. 

In this paper, the 6D $SU(4)$ GHU model on $T^2/Z_2$ is adopted 
and discuss the electroweak symmetry breaking. 
Other than the prediction of realistic weak mixing angle $\sin^2 \theta_W = 1/4$ as mentioned above, 
this model has $Z_2$ and CP symmetries assumed in the ordinary 2HDM 
since the tree level potential is only a quartic term with a gauge coupling constant squared $g^2$ 
originated from the squared commutator in the gauge kinetic term. 
We study the electroweak symmetry breaking in a flat direction 
where the quartic Higgs potential at tree level is vanishing in a vacuum space. 
Then, we calculate one-loop effective potential from the gauge and fermion field contributions. 
As for the fermion field, we discuss that the most dominant contribution to one-loop effective potential 
comes from top quark contribution. 
In GHU, it is known that top quark should be embedded into four-rank totally symmetric tensor representation 
to reproduce top quark mass. 
Therefore, we calculate one-loop effective potential 
by $\overline{{\bf 35}}$ representation of $SU(4)$ as the fermion contribution. 
We will find that our model shows an electroweak symmetry breaking.   
Higgs mass is also calculated and an extension of fermion representation is discussed. 

This paper is organized as follows. 
In section 2, our model is described. 
In section 3, one-loop effective potential of our model is calculated. 
Electroweak symmetry breaking is discussed in section 4. 
The last section is devoted to summary of this paper. 
Appendix A provides a detail calculation of one-loop effective potential 
from the contribution of fields with a typical KK mass spectrum.  
The expression of $SU(4)$ generators are summarized in Appendix B.  
The detail calculation of a general one-loop effective potential for $SU(4)$ $N$-rank totally symmetric representation 
is provided in Appendix C. 


\section{Set up}
We consider a 6D $SU(4)$ GHU model compactified on an orbifold $T^2/Z_2$. 
Let us explain motivations why this model is employed.  
In general, the weak mixing angle $\theta_W$ can be predicted 
if the hypercharge gauge group $U(1)_Y$ in SM is embedded into a simple group. 
In GHU, the minimal simple group is an $SU(3)$ 
 where the SM Higgs doublet belongs to the adjoint representation of $SU(3)$ gauge group. 
However, this model is known to predict a wrong mixing angle $\sin^2\theta_W = 3/4$ 
at the compactification scale \cite{Manton}. 
In \cite{HLM}, it has been discussed that the realistic weak mixing angle 
$\sin^2\theta_W = 1/4~(\sin^2 \theta_W \simeq 0.23$ for an experimental data) 
can be predicted if the SM Higgs doublet belongs to the $SU(3)$ triplet or sextet representation. 
A simplest unitary group including triplet representations under the $SU(3)$ subgroup is an $SU(4)$, 
where the adjoint representation of $SU(4)$ is decomposed as follows, 
${\bf 15} \to {\bf 8} + {\bf 3} + \bar{{\bf 3}} + {\bf 1}$. 
This is why we consider the $SU(4)$ gauge group. 

In GHU, the Higgs doublets are originated from extra spatial components of higher dimensional gauge field. 
In order to realize 2HDM in a context of GHU, at least two extra dimensions are necessary 
 and the simplest 6D case is considered in this paper. 
The number of Higgs doublets in 6D GHU is known to depend nontrivially 
how to compactify the extra two spatial dimensions 
and a 2HDM can be obtained in $T^2/Z_2$ case \cite{SSSW}. 

Now, we discuss our model in detail \cite{HLM}. 
A six-dimensional spacetime is $M^4 \times T^2 / Z_2$, 
 where $M^4$ is a Minkowski spacetime and 
 $T^2/Z_2$ orbifold is a two-dimensional square torus with a radius $R$ divided by $Z_2$ parity. 
By using the $Z_2$ orbifolding, 
we can construct a chiral theory in the four-dimensional theory and also realize a gauge symmetry breaking explicitly.
Our six-dimensional Lagrangian is given by
	\begin{equation}\label{gauge kinetic}
		\mathcal{L}_6 = - \frac{1}{2} \text{Tr} [ F^{MN} F_{MN} ]
		 - \bar{\Psi} i \Gamma^M D_M \Psi  \qquad(M, N = 0, 1, 2, 3, 5, 6) , 
	\end{equation}
where we denote $\Psi$ as a 6D Weyl fermion and 
$\Gamma_M$ are six-dimensional gamma matrices. 
The field strength tensor is defined by
	\begin{align}
		F_{MN}^a  & = \partial_M A_N^a - \partial_N A_M^a - ig [ A_M, A_N ]^a	. 
	\end{align}
The metric convention $\eta_{MN} = \text{diag} ( -1, +1, \cdots , +1 )$ is employed.  
The index $a$ labels $SU(4)$ gauge degrees of freedom.
$g$ is a 6D gauge coupling with mass dimension $-1$.   
The extra dimensional coordinates are represented by $x^5,\: x^6 \in [ 0, \: 2 \pi R ]$.
The identifications for $T^2: x_5 \rightarrow x_5 + 2\pi m R, 
x_6 \rightarrow x_6 + 2\pi n R~(m,n \in \mathbb{Z})$ 
and $Z_2$ parity ($x_{5, 6} \rightarrow - x_{5, 6}$) 
gives the fundamental area of extra dimensions as $x_5 \in [0, \pi R], x_6 \in [0, 2\pi R]$. 

Since the extra dimensions are compactified,  
the boundary conditions on $T^2$ and $Z_2$ parities around fixed points 
which are not transformed by $Z_2$, have to be imposed. 
The properties of the boundary conditions which should be satisfied on $T^2/Z_2$ have been explored in detail in \cite{HNT}. 

The boundary conditions are defined for $T^2$, 
\begin{align}
A_M(x^{\mu}, x_5 + 2 \pi R, x_6 + 2\pi R) = U_{5,6} A_M(x^{\mu}, x_5, x_6) U_{5,6}^\dag,
\end{align}
where $U_{5,6}$ denote gauge degree of freedom associated with the gauge invariance 
and satisfies $[U_5, U_6]=0$ \cite{HNT}. 

We choose the parity matrices $P_i~(i = 0, 1, 2, 3)$ at each fixed point 
$( x^5_i, x^6_i ) =$ (0,0), $(\pi R, 0), (0, \pi R), ( \pi R, \pi R )$ respectively, 
	\begin{align}
			&P_0 = P_1 = \text{diag}( 1, 1, 1, -1 ), 	\\
			&P_2 = P_3 = \text{diag}( 1, 1, -1, -1 ). 
	\end{align}
Note that these parities satisfy $P_3 = P_1 P_0 P_2 = P_2 P_0 P_1$ \cite{HNT}.  
The boundary conditions for the 6D gauge fields at fixed points are given by 
	\begin{align}
	\label{1parity}
			A_{\mu}( x^{\mu}, x^5_i - x^5, x^6_i - x^6 ) &= + P_i A_{\mu}( x^{\mu}, x^5_i + x^5, x^6_i + x^6 ) P^{-1}_i,	\\
			A_{5, 6}( x^{\mu}, x^5_i - x^5, x^6_i - x^6 ) &= - P_i A_{5,6}( x^{\mu}, x^5_i + x^5, x^6_i + x^6 ) P^{-1}_i.
	\end{align}
$Z_2$ parity of $A_\mu$ is chosen by hand to realize a symmetry breaking 
$SU(4) \to SU(2) \times U(1)_Y \times U(1)_X$,  
that of $A_{5,6}$ is fixed by gauge covariance.  
Combining two $Z_2$ parity transformations by $P_{0,1,2}$ as
	\begin{equation}\label{Parity}
		P = \text{diag}\big( (+++), (+++), (++-), (---) \big), 
	\end{equation}
where $\pm$ in the left, center, right parts correspond to $Z_2$ parities $\pm 1$ of $P_{0,1,2}$, respectively. 
Using this expression, the $Z_2$ parities of each component of the gauge fields are obtained
	\begin{align}
		A_\mu &\rightarrow 
			\begin{pmatrix}\mqty{
				(+++) (+++) (++-) (---)\\
				(+++) (+++) (++-) (---)\\
				(++-) (++-) (+++) (--+)\\
				(---) (---) (--+) (+++)
			}\end{pmatrix}, \\
		A_{5, 6} &\rightarrow 
			\begin{pmatrix}\mqty{
				(---) (---) (--+) (+++)\\
				(---) (---) (--+) (+++)\\
				(--+) (--+) (---) (++-)\\
				(+++) (+++) (++-) (---)
			}\end{pmatrix}.
	\end{align}
Taking into account these $Z_2$ parities as well as the periodic boundary conditions on $T^2$, 
 the gauge field can be expanded in terms of mode functions \cite{HNT},  
	\begin{align}
	A_M( x^{M} ) =
		\begin{cases}
		\frac{1}{\sqrt{2} \pi R} A^{( 0, 0 )}_M( x^{\mu} ) 
		\left(
		\begin{array}{c}
		1 \\
		0\\
		\end{array}
		\right)
		+ \frac{1}{\pi R} {\displaystyle \sum_{( n_1, n_2 ) \ne ( 0, 0 )}^\infty} 
		 A^{( n_1, n_2 )}_M( x^{\mu} )
		 \left(
		\begin{array}{c}
		\cos \\
		\sin \\
		\end{array}
		\right)
		 \left( \frac{n_1 x^5 + n_2 x^6}{R} \right) 
		\\
\hspace*{90mm}		{\rm for}~
		\left\{
		\begin{array}{c}
		(+++) \\
		(---) \\
		\end{array}
		\right.,
		\\
		\frac{1}{\pi R}{\displaystyle \sum_{n_1=-\infty}^\infty} 
		{\displaystyle \sum_{n_2 = 0}^\infty} 
		A^{( n_1, n_2 )}_M( x^{\mu} )		
		\left(
		\begin{array}{c}
		\cos \\
		\sin \\
		\end{array}
		\right)
		\left( \frac{n_1 x^5 + (n_2+\frac{1}{2}) x^6}{R} \right)~{\rm for}~
		\left\{
		\begin{array}{c}
		(++-) \\
		(--+) \\
		\end{array}
		\right.,
		\\		
		\frac{1}{\pi R}{\displaystyle \sum_{n_1=0}^\infty} 
		{\displaystyle \sum_{n_2 = -\infty}^\infty} 
		A^{( n_1, n_2 )}_M( x^{\mu} )		
		\left(
		\begin{array}{c}
		\cos \\
		\sin \\
		\end{array}
		\right)
		\left( \frac{(n_1+\frac{1}{2}) x^5 + n_2 x^6}{R} \right)~{\rm for}~
		\left\{
		\begin{array}{c}
		(+-+) \\
		(-+-) \\
		\end{array}
		\right.,
		\\					
		\frac{1}{\pi R}{\displaystyle \sum_{n_1=-\infty}^\infty} 
		{\displaystyle \sum_{n_2 = 0}^\infty} 
		A^{( n_1, n_2 )}_M( x^{\mu} )		
		\left(
		\begin{array}{c}
		\cos \\
		\sin \\
		\end{array}
		\right)
		\left( \frac{(n_1+\frac{1}{2}) x^5 + (n_2+\frac{1}{2}) x^6}{R} \right)~{\rm for}~
		\left\{
		\begin{array}{c}
		(+--) \\
		(-++) \\
		\end{array}
		\right..
		\\				
		\end{cases}
	\label{modes}
	\end{align}
Here we note 
\begin{align}
\sum_{( n_1, n_2 )\neq ( 0, 0 )}^\infty f( n_1, n_2 ) 
= \sum_{n_1=0}^\infty f( n_1, 0) + \sum_{n_1=-\infty}^\infty \sum_{n_2=1}^\infty f( n_1, n_2 ).  
\end{align}

Since the zero modes corresponding to 4D fields appear only in the $(+++)$ part, 
we find that the zero modes of the 4D gauge fields are $SU(2)_L \cross U(1)_Y \cross U(1)_X $ gauge fields, 
and those appear in the off-diagonal components of the extra components of the gauge fields $A_{5,6}$ 
are regarded as two SM Higgs doublets $H_{1,2}$, 
	\begin{align}\label{defHiggs}
		A_{5, 6}^{( 0, 0 )} = \frac{1}{\sqrt{2}}
			\begin{pmatrix}\mqty{
				0 & 0 & 0 & \phi_{1, 2}^{+} \\
				0 & 0 &  0 & \phi_{1, 2}^0  \\ 
				0&0&0&0\\
				\phi_{1, 2}^{-} & \phi_{1, 2}^{0*} & 0 & 0
			}\end{pmatrix},	\quad	
		H_{1, 2}=
			\begin{pmatrix}\mqty{
				\phi_{1, 2}^{+}\\
				\phi_{1, 2}^0
			}\end{pmatrix}.
	\end{align}
As mentioned above, we can see that the Higgs doublets are belonging 
to the triplet of $SU(3)$ being a subgroup of $SU(4)$. 
Therefore, the weak mixing angle in our model is $\sin^2 \theta_W = 1/4$ at the compactification scale. 

Extracting only the Higgs potential at tree level from $(\ref{gauge kinetic})$, we obtain
	\begin{align}
		\text{Tr}[ F_{M N} F^{M N} ]
		&\supset ( - i g )^2 \text{Tr} \left[ A_5, A_6 \right]^2.   
	\end{align}
Note that only the quartic term of Higgs potential is present at tree level. 
Higgs mass terms are forbidden by gauge invariance. 
Rewriting the tree level potential in terms of the two Higgs doublets, we have
	\begin{equation}
		V_{\rm{tree}}( H_1, H_2 ) = - g^2 \text{Tr}\left(\left[ A_5^{( 0, 0 )}, A_6^{( 0, 0 )} \right]^2 \right).
	\end{equation}
We emphasize here a remarkable property of GHU. 
The coupling constant of Higgs quartic terms is the gauge coupling squared similar to SUSY case. 
This means that the predictability of physics in the Higgs sector is not only enhanced, 
but also CP invariance assumed in the ordinary 2HDM is trivially satisfied since the gauge coupling is real. 

The commutator of $A^{( 0, 0 )}_5$ and $A^{( 0, 0 )}_6$ is computed as 
	\begin{align}
		 \comm{A^{( 0, 0 )}_5}{A^{( 0, 0 )}_6} & = \frac{1}{2}
 			\begin{pmatrix}\mqty{
			H_1H_2^{\dagger} - H_2 H_1^{\dagger} & 0 &0 \\
			0 & 0 & 0\\
			0& 0 & H_1^{\dagger}H_2 - H_2^{\dagger}H_1 
			}\end{pmatrix}.
	\end{align}
Thus, the Higgs potential at the tree level in this model is
	\begin{equation}
		V_{\rm{tree}}( H_1, H_2 ) = \frac{g^2}{2} \big\{ ( H_1^{\dagger} H_1 ) ( H_2^{\dagger} H_2 ) 
		+ ( H_1^{\dagger} H_2 ) ( H_2^{\dagger} H_1 ) - ( H_2^{\dagger} H_1 )^2 - ( H_1^{\dagger} H_2 )^2 \big\}.
	\end{equation}
Note that the above Higgs potential has a $Z_2$ symmetry $(H_1, H_2) \to (H_1, -H_2)$. 
In the ordinary 2HDM, a softly broken $Z_2$ symmetry is assumed to forbid FCNC. 
In our model, such $Z_2$ symmetry is incorporated at tree level and 
 we will see that soft breaking mass is generated at one-loop.  

Suppose that two Higgs fields have the following vacuum expectation values (VEV),
	\begin{align}
		\ev{H_1} =
			\begin{pmatrix} \mqty{
			0 \\
			\frac{v_1}{\sqrt{2}} 
			}\end{pmatrix},
		\:
		\ev{H_2} = 
			\begin{pmatrix} \mqty{
			0 \\ 
			\frac{v_2}{\sqrt{2}} 
			}\end{pmatrix} 
		\qquad( v_1, v_2 \in \mathbb{R} ),
	\end{align}
which corresponds to the flat direction of the potential $ V_{\text{tree}}( v_1, v_2 ) = 0 $ 
 similar to the D-flat direction in SUSY theory. 
This is a minimum of the potential at tree level and the VEV $v_{1,2}$ are undetermined. 
In order to fix the magnitude of the VEV, we have to calculate one-loop effective potential of our model. 

\section{One-loop effective potential}
In this section, one-loop effective potential of our model will be calculated. 
The formula of effective potential is given by
	\begin{align}
		V_{\text{eff}}^{\text{1-loop}} = \frac{(-1)^{F}}{2} N \sum_{n_1=0}^\infty \sum_{n_2=0}^\infty 
		\int \frac{\dd^4 p}{( 2 \pi )^4} \log( p^2 + M^2_{n_1, n_2}),
	\end{align}
where the sum is taken for all KK modes with a mass $M_{n_1, n_2}~(n_{1,2} =0$ and $\mathbb{N})$ 
running in the loop diagram. 
$F$ represents a fermion number, namely $F=0$ for bosons and $F=1$ for fermions. 
$N$ is a number of degrees of freedom of the fields running in the loop. 
If the two Higgs fields have VEV $v_1, v_2$, the typical KK mode mass takes a form (which will be derived later) 
	\begin{align}
		M_{n_1, n_2}^2
		& = \left(\frac{n_1\pm \alpha_1}{R}\right)^2 
		+\left(\frac{n_2\pm \alpha_2}{R}\right)^2 \qquad (\alpha_i = g_4 v_i R) ,
	\end{align}
where $g_4$ is a 4D gauge coupling constant related to 6D gauge coupling as $g = {\sqrt{2} \pi R}{g_4}$. 
$\alpha_i$ are dimensionless parameters of the Higgs field VEV.  

Then, the contribution to the one-loop effective potential by the field with the above KK mass is obtained 
	\begin{align}
		V_{\text{eff}}^{\text{1-loop}}
		&= - \frac{(-1)^F }{16 \pi^7 R^4} \sum_{( k_1, k_2 ) \neq ( 0, 0 )}
		\frac{e^{2 \pi i ( k_1 \alpha_1 + k_2 \alpha_2 )}}{(k_1^2 + k_2^2)^3}	\notag \\
		&= - \frac{(-1)^F}{16 \pi^7 R^4} \sum_{( k_1, k_2 ) \neq ( 0, 0 )}
		\frac{\cos( 2 \pi k_1 \alpha_1) \cos( 2 \pi k_2 \alpha_2)}{(k_1^2 + k_2^2)^3}\notag \\
		&\equiv - \frac{(-1)^F}{16 \pi^7 R^4} \sum_{( k_1, k_2 ) \neq ( 0, 0 )} \nu(\alpha_1, \alpha_2).
	\end{align}
The detail derivation of the effective potential is described in Appendix A.

\subsection{$SU(4)$ gauge fields}
To calculate the one-loop effective potential, we only need to find the KK mass spectrum of the fields.
We look at the gauge kinetic term,
	\begin{align}
		F^a_{\mu 5}F^{a \mu 5} \supset D_5 A^a_{\mu} D^5 A^{a \mu} 
		= A^c_{\mu} [ - ( D_5^{cc^{\prime}} )^2 ]A^{c^{\prime}}_{\mu}.
	\end{align}
Thus, KK mass is obtained by covariant derivative in the extra dimension.
	\begin{align}
		\hat{M}^2 = - ( D^2_5 + D^2_6 ), \qquad 
		( D^{ab}_{5,6} = \delta^{a b} \partial_{5, 6} + i g f^{acb} A^c_{5, 6} ),
	\end{align}
where $f^{abc}$ is a structure constant of $SU(4)$. 
Now, we can obtain KK mass of the gauge fields in an $SU(4)$ adjoint representation. 
In order to obtain the mass matrix, 
we need information of 15 generators $T^a = t^a / 2$ of $SU(4)$ summarized in Appendix B. 
Since the VEVs of Higgs fields are assumed in the 11 direction of the $SU(4)$ generator from (\ref{defHiggs}) 
	\begin{align}
		\ev{A_{5 ,6}^{11}} = \frac{v_{1, 2}}{\sqrt{2}},
	\end{align}
the nonvanishing structure constants $f^{a\, 11\, b}$ are found as
	\begin{align}
		\begin{cases}
			f^{1\, 11\, 10} = f^{6\, 11\, 14} = f^{7\, 11\, 13}=\displaystyle\frac{1}{2}, 	\\[3mm]
			f^{2\, 11\, 9} = f^{3\, 11\, 12}= \displaystyle-\frac{1}{2},	\\[3mm]
			f^{8\, 11\, 12}=\displaystyle\frac{\sqrt{3}}{6},  f^{15\, 11\, 12}=\sqrt{\frac{2}{3}}.
		\end{cases}
	\end{align}
Using the mode expansions (\ref{modes}) and diagonalizing the mass matrix with these structure constants, 
we obtain mass eigenvalues,
	\begin{align}
		M^2_{n_1, n_2} &= \frac{\tilde{n}_1^2 + \tilde{n}_2^2}{R} \times 3,\:  
		\frac{n_1^2 + (n_2 + \frac12)^2}{R} \times 2,\: 
		\frac{(n_1 \pm \frac{\alpha_1}2 )^2 + (n_2 + \frac{1}{2} \pm \frac{\alpha_2}2 )^2}{R} \times 2,\: 
		\nonumber \\
		& \quad 
		\frac{(\tilde{n}_1 \pm \frac{\alpha_1}2 )^2 + (\tilde{n}_2 \pm \frac{\alpha_2}2 )^2}{R} \times 2,\: 
		\frac{(\tilde{n}_1 \pm \alpha_1 )^2 + (\tilde{n}_2 \pm \alpha_2 )^2}{R}, \\
		&(n_1, n_2) = (0 \cdots \infty, -\infty \cdots \infty), \notag \\
		&(\tilde{n}_1, \tilde{n}_2) = (0, 1 \cdots \infty), (1 \cdots \infty, -\infty \cdots \infty), 
	\end{align}
where the degeneracy is explicitly specified for each eigenvalue. 
Then, this gives all of 15 gauge field KK masses. 
These mass eigenvalues can be further combined 
if we consider the ranges of $(n_1, n_2)$ and $(\tilde{n}_1, \tilde{n}_2)$ carefully. 
	\begin{align}
		M^2_{n_1, n_2} &= 
		\frac{(n_1 + \frac{\alpha_1}2 )^2 + (n_2 + \frac{1}{2} + \frac{\alpha_2}2 )^2}{R} \times 2,\: 
		 \quad 
		\frac{(n_1 + \frac{\alpha_1}2 )^2 + (n_2 + \frac{\alpha_2}2 )^2}{R} \times 2,\: \notag\\
		& \frac{(n_1 + \alpha_1 )^2 + (n_2 + \alpha_2 )^2}{R}~(-\infty < n_{1,2} < \infty). 
	\end{align}
  
\noindent
Since we are interested in determining the magnitude of Higgs VEV from the one-loop effective potential, 
the masses independent of Higgs VEV are simply ignored.


Finally, we obtain one-loop effective potential for an $SU(4)$ gauge field. 
	\begin{align}
		V^g_{\text{eff}}
		&=\frac{4}{2} \sum_{n_1,n_2=-\infty}^\infty \int \frac{\dd^4 p}{(2 \pi)^4}\ 
		\left\{
		2 \log \left[p^2 + \frac{(n_1+\frac{\alpha_1}{2})^2+(n_2 + \frac{1}{2} + \frac{\alpha_2}{2})^2}{R^2}\right]
		\right. \notag \\
		&\quad \left.  +2 \log \left[p^2 + \frac{(n_1+\frac{\alpha_1}{2})^2+(n_2+\frac{\alpha_2}{2})^2}{R^2}\right]		
				+ \log \left[p^2 +\frac{(n_1+{\alpha_1})^2+(n_2+{\alpha_2})^2}{R^2}\right] \right\} \notag \\
		& = - \frac{1}{4 \pi^7 R^4} \sum_{( k_1, k_2 ) \neq ( 0, 0 )} 
		\left \{ 2 (1 + (-1)^{k_2}) \nu \left( \frac{\alpha_1}{2}, \frac{\alpha_2}{2} \right) + \nu( \alpha_1, \alpha_2 )\right \},
	\end{align}
where the expression of the one-loop effective potential for the gauge field with anti-periodic boundary conditions is used. 
	\begin{align}
		\nu \left(\alpha_1, \alpha_2 + \frac{1}{2}\right)
		&=  \frac{\cos( 2 \pi k_1 \alpha_1) \cos( 2 \pi k_2 (\alpha_2 + \frac{1}{2}))}{(k_1^2 + k_2^2)^3} \notag \\
		&=(-1)^{k_2} \frac{\cos( 2 \pi k_1 \alpha_1) \cos( 2 \pi k_2 \alpha_2)}{(k_1^2 + k_2^2)^3} \notag \\
		&=(-1)^{k_2}\nu(\alpha_1, \alpha_2). 
	\end{align}

\subsection{Fermions}
In this section, we calculate one-loop effective potential from fermion loop. 
If the SM fermions are embedded in 6D fermions, 
Yukawa interaction is generated from the gauge interaction $g \overline{\psi} A_{5,6} \psi$ in GHU. 
4D effective Yukawa coupling is obtained by an overlap integral of zero mode functions of SM fermions. 
For fermions except for top quark, fermion masses are typically given by $M_W e^{-MR}$, 
where $M_W, M$ are weak scale, 6D fermion mass (referred as bulk mass). 
For top quark case, top quark mass can be realized 
if top quark is embedded into a 6D massless fermion in the four-rank totally symmetric tensor of the gauge group 
\cite{SSS,CCP}. 
On the other hand, one-loop effective potential from massive fermion contributions 
is known to be suppressed by a Boltzmann-like factor $e^{-\pi MR}$. 
Therefore, the most dominant contribution to one-loop effective potential is given by top quark, 
which will be calculated in detail. 

\subsubsection{fundamental representation fermions}
Although we would like to calculate one-loop effective potential from top quark obtained 
 from four-rank totally symmetric tensor of $SU(4)$, 
 we first consider the $SU(4)$ fundamental representation fermions as 6D Weyl fermion 
 to understand a structure of KK mass spectrum, 
	\begin{align}
		\Psi=
			\begin{pmatrix}\mqty{
			\psi_1 \\
			\psi_2 \\
			\psi_3 \\
			\psi_4 
			}\end{pmatrix}.
	\end{align}
$Z_2$ parity of $\Psi$ is defined as 
	\begin{align}
		\Psi(-x^5, -x^6)
		&=P (-i \Gamma_5 \Gamma_6)\Psi(x^5, x^6) \notag \\
		&=-P(I_4 \otimes \sigma_3) \Psi(x^5, x^6),
	\end{align}
where 6D gamma matrices are given
	\begin{align}
		\Gamma^{\mu}=
			\begin{pmatrix}\dmat{
			\gamma^{\mu},\gamma^{\mu}
			}\end{pmatrix}, \quad 
		\Gamma^{5}=
			\begin{pmatrix}\mqty{
			 & i \gamma^{5}\\ i \gamma^{5}& 
			}\end{pmatrix}, \quad
		\Gamma^{6}=\begin{pmatrix}\mqty{
			 & \gamma^{5}\\ - \gamma^{5}& 
			}\end{pmatrix}
		\qquad (\mu = 0, 1, 2, 3)
	\end{align}
and 6D chiral operator is defined as
	\begin{align}
		\Gamma^7
		&=\Gamma^0\Gamma^1\Gamma^2\Gamma^3\Gamma^5\Gamma^6
		=\begin{pmatrix}\dmat{
			-\gamma^5, \gamma^5
			}\end{pmatrix}.
	\end{align}	
$\gamma^\mu, \gamma^5$ are 4D gamma matrices. 
Then, the parity assignments are explicitly shown in components 
	\begin{align}
		\Psi=\begin{pmatrix}\mqty{
		\psi^{(+++)}_{1L}+\psi^{(---)}_{1R}\\[2mm]
		\psi^{(+++)}_{2L}+\psi^{(---)}_{2R}\\[2mm]
		\psi^{(++-)}_{3L}+\psi^{(--+)}_{3R}\\[2mm]
		\psi^{(---)}_{4L}+\psi^{(+++)}_{4R}
		}\end{pmatrix}.
	\end{align}
Since Lagrangian for $\Psi$ is 
\begin{align}
{\cal L} = \overline{\Psi} (i \Gamma^\mu D_\mu + i \Gamma^5 D_5 + i \Gamma^6 D_6) \Psi, 
\end{align}
where the covariant derivative is $D_M = \partial_M -ig A_M$, 
the mass squared matrix is found 
	\begin{align}
		\tilde{M}^2_{n_1,n_2}
		&=(D_5\Gamma^5+D_6\Gamma^6)^2\notag \\
		&=M_{n_1,n_2}^2+\frac{1}{2}\Gamma^5\Gamma^6\comm{D_5}{D_6}\notag \\
		&=M_{n_1,n_2}^2 + i g^2 ( I_4\otimes \sigma_3)\comm{A_5}{A_6}. 
	\end{align}
Noting that we consider the flat direction $[A_5, A_6]=0$, 
 the mass squared matrix of fermions takes the same form as that of $SU(4)$ gauge fields. 
More precisely, the mass squared mass matrix of fermion in the $SU(4)$ fundamental representation 
is the same form of $SU(2)$ doublet in the decomposition of $SU(4)$ gauge field.  
This is because the decomposition of the $SU(4)$ fundamental representation into $SU(2)$ representation as follows. 
	\begin{align}
		\ytableausetup{mathmode, boxsize=2em}
			\begin{ytableau}
			4 
			\end{ytableau}
		&\xRightarrow{SU(3)} 
			\begin{ytableau}
			3
			\end{ytableau}
		+\textbf{1}_{(++-)} \notag \\
		& \xRightarrow{SU(2)} 
			\begin{ytableau}
			2
			\end{ytableau}
		+\textbf{1}_{(---)}+\textbf{1}_{(++-)}.
	\end{align}
Since only the $SU(2)$ doublet field interacts with Higgs field, 
the KK spectrum of fermion in $SU(4)$ fundamental representation is essentially 
the same as that of $SU(2)$ doublet fermion, 
which is also the same as that of $SU(2)$ doublet components of $SU(4)$ gauge field. 
Note also that an $SU(2)$ singlet field does not interact with Higgs field 
 and gives just constant in one-loop effective potential. 
Therefore, one-loop effective potential of fermion can be calculated 
by using information of the representation after $SU(2)$ decomposition.

\subsubsection{One-loop effective potential from top quark}

As mentioned above, top quark contribution to one-loop effective potential is the most dominant one of fermions 
 since the vanishing bulk mass is required to reproduce top quark mass 
 and this leads to no suppression via bulk mass to one-loop effective potential. 
Furthermore, if we suppose that  top quark is realized as a zero mode of the 6D fermion field, 
we have to embed it into the fermion in a higher rank representation 
since top quark mass becomes a weak boson mass in case of the fundamental representation. 
It is known that GHU can successfully reproduce the top quark mass, 
especially when the top quark is embedded in a four-rank totally symmetric representation \cite{SSS, CCP}. 
Therefore, we first consider the $\overline{\textbf{35}}$ representation, 
which is the four-rank totally symmetric representation of $SU(4)$ 
expressed by Young tableau as \ytableausetup{smalltableaux}
	\begin{ytableau}
		4&5&6&7
	\end{ytableau}.

To find the KK mass spectrum of the fermion in $\overline{\textbf{35}}$ representation, 
we need to know the form of covariant derivative depending on the representation. 
The covariant derivative for a $\overline{\textbf{35}}$ representation fermion is found 
	\begin{equation}
		D_{5,6}\Psi_{ijkl}
		=(\partial_{5,6}\delta_{ii^{\prime}}+4igA^a_{5,6} T^a_{ii^{\prime}})\Psi_{i^{\prime}jkl}, 
	\label{cov35}
	\end{equation}
which can be further extended for $N$-rank totally symmetric tensor representation
	\begin{equation}\label{fermioncdv}
		D_{5,6}\Psi_{i_1 \dots i_N }
		=(\partial_{5,6}\delta_{i_1 i_1^{\prime}}+NigA^a_{5,6} T^a_{i_1 i^{\prime}_1})\Psi_{i^{\prime}_1 \dots i_N }.
	\end{equation}
Namely, the number of indices of representation is multiplied in front of the generator in the covariant derivative. 
This is reflected as a factor of the mass splitting in the KK spectrum after developing Higgs VEV. 
Now, we decompose the $\overline{\textbf{35}}$ representation of $SU(4)$ 
into $SU(3)$ and $SU(2)$ representations.
	\begin{align}
		\ytableausetup{mathmode, boxsize=normal}
		\ydiagram{4}
		&\xRightarrow{SU(3)}
		\ydiagram{4}_{(+++)}+\ydiagram{3}_{(++-)}+\ydiagram{2}_{(+++)}+\ydiagram{1}_{(++-)}+\textbf{1}_{(+++)}\notag \\[2mm]
		&\xRightarrow{SU(2)}
		\ydiagram{4}_{(+++)}+\ydiagram{3}_{(---)}+\ydiagram{3}_{(++-)}\notag \\
		&\qquad \qquad +2\ \ydiagram{2}_{(+++)}+\ydiagram{2}_{(--+)}+2\ \ydiagram{1}_{(---)}
		+2\ \ydiagram{1}_{(++-)}+ 5\ \text{singlet}. 
	\label{35decompose}
	\end{align}
We impose the $Z_2$ parity on a fermion in $\overline{{\bf 35}}$ representation as
\begin{align}
\Psi_{ijkl}(-x^5, -x^6) = - {\cal R}_{\overline{{\bf 35}}}(P) \Psi_{ijkl}(x^5, x^6). 
\end{align}
In the above decomposition, top quark is included as a zero mode 
in $\overline{{\bf 15}}$ representation of $SU(3)$ \cite{CCP}. 
More precisely, top quark corresponds to zero modes of ${\bf 2}_L$ and ${\bf 1}_R$ representations 
in the $SU(2)$ decomposition of $\overline{\textbf{15}}$.  
Note here that many massless exotic fermions appear from $\overline{{\bf 15}}$, 
$\overline{{\bf 10}}$, $\overline{{\bf 6}}$ and $\overline{{\bf 1}}$ representations of $SU(3)$. 
Of these massless exotic fermions, those from $SU(2)$ singlet $\overline{{\bf 1}}$ give just constant contributions 
to one-loop effective potential since those are independent of Higgs VEV. 
Therefore, these are harmless. 
For other massless exotic fermions, they can be made massive  
by introducing 4D Weyl fermions with the opposite chirality and conjugate representation 
on the brane at the fixed point and making Dirac mass terms with the massless exotic fermions. 
The mass of the exotic fermions obtained through the Dirac mass terms is also independent of Higgs VEV, 
therefore it is not necessary for one-loop effective potential to take into account their contributions.     

From information (\ref{cov35}) and (\ref{35decompose}), 
the KK mass spectrum of fermion in the $\overline{{\bf 35}}$ representation is obtained 
in the same way as the gauge field.
	\begin{align*}
		\ytableausetup{mathmode, boxsize=normal}
		\ydiagram{4}_{(+++)} &\rightarrow \frac{(n_1 \pm 2 \alpha_1)^2 
		+ (n_2 \pm 2 \alpha_2)^2 }{R^2}, \frac{(n_1 \pm \alpha_1)^2 
		+ (n_2 \pm \alpha_2)^2 }{R^2}, \frac{n_1^2 + n_2^2 }{R^2}, \\
		\ydiagram{3}_{(---)} &\rightarrow \frac{(n_1 \pm \frac{3}{2} \alpha_1)^2 
		+ (n_2 \pm \frac{3}{2} \alpha_2)^2 }{R^2},  \frac{(n_1 \pm \frac{1}{2} \alpha_1)^2 
		+ (n_2 \pm \frac{1}{2} \alpha_2)^2 }{R^2}, \\
		\ydiagram{3}_{(++-)} &\rightarrow \frac{(n_1 \pm \frac{3}{2} \alpha_1)^2 
		+ (n_2 + \frac{1}{2} \pm \frac{3}{2} \alpha_2)^2 }{R^2},  
		\frac{(n_1 \pm \frac{1}{2} \alpha_1)^2 + (n_2 + \frac{1}{2} \pm \frac{1}{2} \alpha_2)^2 }{R^2}, \\ 
		\ydiagram{2}_{(--+)}& \rightarrow \frac{(n_1 \pm \alpha_1)^2 
		+ (n_2 \pm \alpha_2)^2 }{R^2}, \frac{n_1^2 + n_2^2 }{R^2}, \\
		\ydiagram{2}_{(---)}& \rightarrow \frac{(n_1 \pm \alpha_1)^2 
		+ (n_2 + \frac{1}{2} \pm \alpha_2)^2 }{R^2}, \frac{n_1^2 + (n_2 + \frac{1}{2})^2}{R^2}, \\
		\ydiagram{1}_{(---)}& \rightarrow \frac{(n_1 \pm \frac{1}{2} \alpha_1)^2 
		+ (n_2 \pm \frac{1}{2} \alpha_2)^2 }{R^2}, \\
		\ydiagram{1}_{(++-)}& \rightarrow \frac{(n_1 \pm \frac{1}{2} \alpha_1)^2 
		+ (n_2 + \frac{1}{2} \pm \frac{1}{2} \alpha_2)^2 }{R^2}. 
	\end{align*}
%
We can read off the one-loop effective potential for top quark from the above KK spectrum
	\begin{align}
		V_{\text{eff}}^{f}
		=\frac{3}{4 \pi^7 R^4} &\sum_{( k_1, k_2 ) \neq ( 0, 0 )} 
		\bigg[ \nu(2 \alpha_1, 2 \alpha_2) + \{ 1 + (-1)^{k_2} \} \, \nu \left( \frac{3}{2} \alpha_1, 
		\frac{3}{2} \alpha_2 \right) \notag \\
		 &\hspace{40pt}  + \{ 3 + (-1)^{k_2} \} \, \nu(\alpha_1, \alpha_2) + \{ 3 + 3 (-1)^{k_2} \} \, \nu 
		 \left( \frac{1}{2}\alpha_1, \frac{1}{2}\alpha_2 \right) \bigg],
	\end{align}
where we note that a color factor 3 is multiplied.  
\section{Electroweak symmetry breaking}

\begin{figure}[H]
		\centering
		\includegraphics[width=100mm]{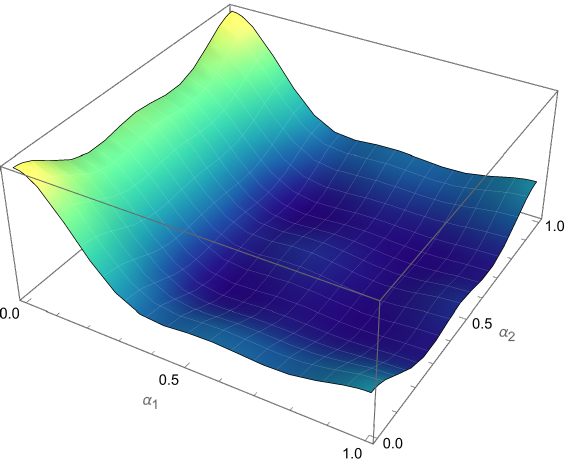}
		\caption{$V^{\text{1-loop}}_{\text{eff}}$ by $SU(4)$ gauge fields and $\overline{{\bf 35}}$ representation fermion}
		\label{fig:potentialplot}
	\end{figure}

In this section, we discuss whether the electroweak symmetry breaking will take place in our 2HDM. 
The effective potential of Higgs fields were found in previous sections. 
	\begin{align}
		V^{\text{1-loop}}_{\text{eff}} 
		&= V^{g}_{\text{eff}}+V^{f}_{\text{eff}} \notag \\
		&=\frac{1}{4 \pi^7 R^4} \sum_{( k_1, k_2 ) \neq ( 0, 0 )} 
		\bigg[ 
		3 \nu(2 \alpha_1, 2 \alpha_2) + 3 \{ 1 + (-1)^{k_2} \} \, 
		\nu \left( \frac{3}{2} \alpha_1, \frac{3}{2} \alpha_2 \right) \notag \\
		 &\hspace{80pt}  + \{ 8 + 3(-1)^{k_2} \} \, \nu(\alpha_1, \alpha_2) 
		 + 7 \{ 1 +  (-1)^{k_2} \} \, \nu \left( \frac{1}{2}\alpha_1, \frac{1}{2}\alpha_2 \right) 
		 \bigg]. 
	\end{align}
The plot of $V^{\text{1-loop}}_{\text{eff}}$ by $SU(4)$ gauge fields 
and a fermion in the $\overline{{\bf  35}}$ representation is shown in Figure \ref{fig:potentialplot}.\\%
The potential minimum is found at $\alpha^{\text{min}}_1=0.438 \dots, \alpha^{\text{min}}_2=0.299 \dots$ 
as can be seen from Figs.~2,3 
and the electroweak symmetry is spontaneously broken.\footnote{In GHU, the electroweak symmetry breaking 
$SU(2)_L \times U(1)_Y \to U(1)_{\rm em}$ is realized at $0 < \alpha_{1,2} < 1$.}
	\begin{figure}[H]
		\begin{tabular}{cc}
			\begin{minipage}{.50 \textwidth}
				\centering
				\includegraphics[width=70mm]{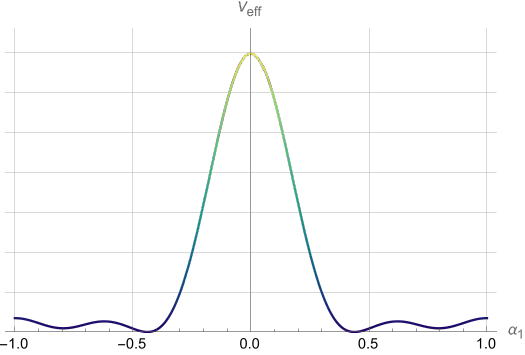}
				\caption{One-loop effective potential in 
					$H_1$ direction}
			\end{minipage}
			\begin{minipage}{.50 \textwidth}
				\centering
				\includegraphics[width=70mm]{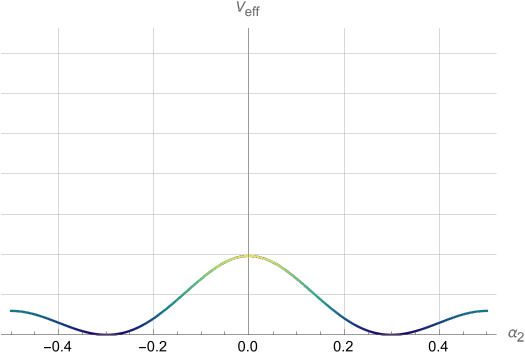}
				\caption{One-loop effective potential 
					in $H_2$ direction}
			\end{minipage}
		\end{tabular}
\label{35alpha}
	\end{figure}
	
Neutral Higgs boson mass matrix elements are obtained 
from the second derivative of one-loop effective potential at the minimum. 
	\begin{align}
		\begin{cases}
			m^2_{11}=( \frac{g_4}{R} )^2 \cross 0.0645 \dots,  \\[2mm]
			m^2_{22}=( \frac{g_4}{R} )^2 \cross 0.0796 \dots,  \\[2mm]
			m^2_{12}=m^2_{21} = -( \frac{g_4}{R} )^2 \cross 0.0109 \dots. 
		\end{cases}
	\end{align}
We note that the 4D gauge coupling and the compactification scale related to the W-boson mass as
	\begin{align}
		m^2_W=\frac{g_4^2}{4}(v_1^2 + v_2^2) 
		=\frac{(\alpha_1^{\text{min}})^2+(\alpha_2^{\text{min}})^2}{4R^2} = (80.4~{\rm GeV})^2,
	\end{align}
where $v=\sqrt{v_1^2+v_2^2} \sim 246$ GeV. 
SM Higgs boson mass corresponds to a smaller eigenvalue of the above $2 \times 2$ neutral Higgs boson mass matrix. 
Unfortunately, we find both compactification scale and Higgs mass to be very small in our model. 
	\begin{align}
		\frac{1}{R} \sim 303 \ \text{GeV}, \quad  
		m_H \sim 48.0 \  \text{GeV},
	\end{align}
where $g_4 = 2m_W/v \sim 0.653$ is used. 

In order to improve this result, the values of the minimum need to be located closer to the origin 
 and to make $\alpha_{1,2}$ small at the order of ${\cal O}(0.01)$. 
Then, the compactification scale is of order of ${\cal O}(10)$ TeV and Higgs mass can be 125 GeV. 
One of the simple extensions is to introduce a fermion including top quark 
in the representation higher than four-rank totally symmetric representation. 
In that case, the minimum of the potential is expected to be closer to the origin 
since the period of one-loop effective potential becomes smaller.  
From this observation, we extended our analysis to a general case for $N$-rank totally symmetric representation 
of $SU(4)$ and analyzed a general form of one-loop effective potential 
described in detail in Appendix C. 
Contrary to the above expectation, we could not enhance enough the compactification scale and Higgs mass 
although the electroweak symmetry breaking still works. 

\section{Summary}
In this paper, we studied the electroweak symmetry breaking in 2HDM based on 
a 6D $SU(4)$ GHU model compactified on $T^2/Z_2$. 
This is one of the simple 2HDM where a realistic weak mixing angle $\sin^2\theta_W= 1/4$ is predicted \cite{HLM}.   

The remarkable properties of this model other than the weak mixing angle prediction are as follows. 
In the ordinary 4D 2HDM, a softly broken $Z_2$ and CP symmetries are assumed in the potential 
to circumvent FCNC at tree level, CP violation. 
Furthermore, the ordinary 4D 2HDM models except for MSSM do not solve the hierarchy problem. 
Therefore, the SM Higgs mass cannot be predicted and is unstable against quantum corrections. 
Since these models also have many free parameters in the Higgs potential, 
they cannot determine whether the electroweak symmetry breaking takes place.  

In our model, on the other hand, $Z_2$ and CP symmetries are incorporated 
since the potential at tree level is generated from the commutator squared in the gauge kinetic term and 
the mass terms to break $Z_2$ symmetry softly are generated at one-loop correction.  
One-loop effective potential is calculable and finite since the Higgs sector is controlled 
by the higher dimensional gauge symmetry. 
Therefore, the possibility of electroweak symmetry breaking and Higgs mass can be predicted. 

We have calculated one-loop effective potential in 6D $SU(4)$ GHU model on $T^2/Z_2$. 
As for the fermion contribution to the potential, 
we calculated the potential of a fermion in $\overline{{\bf 35}}$ 
representation of $SU(4)$, 
which includes top quark and is the most dominant contribution. 
As a result, we have shown that the electroweak symmetry breaking indeed happened. 
However, it turned out that the compactification scale and Higgs mass become too small. 
Since this results are expected to be improved 
if the fermion in a higher-rank totally symmetric representation is introduced, 
we extended our analysis to calculate one-loop effective potential of fermion 
in an $N$-rank totally symmetric representation. 
The results were unfortunately not so improved though the electroweak symmetry breaking still happened.  
 
Although a realistic SM Higgs mass could not be obtained in this paper, 
there is another approach \cite{SSS} to enhance the compactification scale and Higgs mass, 
which is not discussed in this paper. 
It has been known that the compactification scale can be enhanced 
if the localized 4D gauge kinetic terms are introduced at the fixed points. 
It would be very interesting to explore such a possibility and to obtain a realistic SM Higgs mass. 
If a realistic SM Higgs mass can be realized, 
it would be also interesting to find the mass spectrum of other physical Higgs bosons 
and study our predictions for their experimental signatures. 
These issues are left for future works.  \\

\subsection*{Acknowledgments}
This work was supported by JSPS KAKENHI Grant No. JP22KJ2621 (T.H.)
and JST, the establishment of university fellowships towards the creation of science technology innovation, 
Grant Number JPMJFS2138 (K.A.).

\vspace*{10mm}

\appendix
\section{One-loop effective potential}
The formula of effective potential is given by \cite{ABQ}
	\begin{align}
		V_{\text{eff}}^{\text{1-loop}} = \frac{(-1)^{F}}{2} N \sum_{n_1=0}^\infty \sum_{n_2=0}^\infty 
		\int \frac{\dd^4 p}{( 2 \pi )^4} \log( p^2 + M^2_{n_1, n_2}). 
	\end{align}
It can be written in the Schwinger representation, 
	\begin{align}\label{Effective}
		V_{\text{eff}}^{\text{1-loop}}
		&= - \frac{( - 1 )^{F}}{2} \sum_{n_1, n_2=0}^\infty \int_0^\infty 
		\frac{\dd t}{t} \int \frac{d^4 p}{( 2 \pi )^4} e^{- t ( p^2 + M^2_{n_1, n_2}) } \notag \\
		& = - \frac{( - 1 )^{F}}{32 \pi^2} \sum_{n_1, n_2=0}^\infty  \int_0^\infty \frac{\dd t}{t^3} 
		e^{- t M^2_{n_1, n_2}}	\notag \\
		& = - \frac{( - 1 )^{F}}{32 \pi^2} \sum_{n_1, n_2=0}^\infty \int_0^\infty \dd{l} 
		\; le^{-M^2_{n_1, n_2}/l}.
	\end{align}
For the KK mass spectrum
	\begin{align}
		M_{n_1, n_2}^2
		& = \left(\frac{n_1\pm \alpha_1}{R}\right)^2 
		+\left(\frac{n_2\pm \alpha_2}{R}\right)^2, 
	\end{align}
we transform the part of the sum of KK mode as follows
	\begin{align}
\sum_{n_1, n_2=0}^\infty e^{-M^2_{n_1, n_2}/l} 
		&= \sum_{n_1, n_2=0}^\infty \left\{
		\exp[-\left(\frac{(n_1 + \alpha_1)^2}{R^2 l} + \frac{(n_2 + \alpha_2 )^2}{R^2 l}\right)] 
		\right. \notag \\
		& \left. \qquad
		+ \exp[-\left(\frac{(n_1 - \alpha_1)^2}{R^2 l} + \frac{(n_2 - \alpha_2 )^2}{R^2 l}\right)] \right\}	
		\notag \\
		&=\sum_{n_1, n_2=-\infty}^\infty \exp[-\left(\frac{(n_1 + \alpha_1)^2}{R^2 l}
		+\frac{(n_2 + \alpha_2 )^2}{R^2 l}\right)].
	\end{align}
Applying Poisson resummation formula
	\begin{align}
		\sum_{n=-\infty}^\infty \exp[-\frac{(n+\alpha)^2}{R^2 l}]=R \sqrt{l \pi}\sum_{k=-\infty}^\infty 
		\exp[2 \pi i k \alpha - R^2 l \pi^2 k^2].
	\end{align}
to the mode sum, 
the sum with respect to KK momentum is transformed into the sum with respect to winding number of $T^2$.  
Therefore, the contribution to the one-loop effective potential by the field with the above KK mass is obtained 
	\begin{align}
		V_{\text{eff}}^{\text{1-loop}} 
		= - \frac{(-1)^F}{32 \pi^2} \sum_{k_1, k_2=-\infty}^\infty 
		e^{2 \pi i ( k_1 \alpha_1 + k_2 \alpha_2 )} \int_0^\infty \dd{l} \; R^2 l^2 \pi  
		e^{- R^2 l \pi^2 ( k_1^2 + k_2 ^2 ) }.
	\end{align}
Changing the integral variable $l$ to $l^{\prime} =  R^2 l \pi^2 ( k_1^2 + k_2 ^2 ) $ 
and subtract the divergent term from no winding mode $ (k_1, k_2) = ( 0, 0 )$, we have
	\begin{align}
		V_{\text{eff}}^{\text{1-loop}}
		&= - \frac{(-1)^F }{16 \pi^7 R^4} \sum_{( k_1, k_2 ) \neq ( 0, 0 )}
		\frac{e^{2 \pi i ( k_1 \alpha_1 + k_2 \alpha_2 )}}{(k_1^2 + k_2^2)^3}	\notag \\
		&= - \frac{(-1)^F}{16 \pi^7 R^4} \sum_{( k_1, k_2 ) \neq ( 0, 0 )}
		\frac{\cos( 2 \pi k_1 \alpha_1) \cos( 2 \pi k_2 \alpha_2)}{(k_1^2 + k_2^2)^3}\notag \\
		&\equiv - \frac{(-1)^F}{16 \pi^7 R^4} \sum_{( k_1, k_2 ) \neq ( 0, 0 )} \nu(\alpha_1, \alpha_2).
	\end{align}
In this way, we obtain the finite one-loop effective potential.

\section{\mbox{\boldmath $SU(4)$} generators}
$SU(4)$ generators are summarized as follows. 
	\begin{align*}
		t^1&=
			\begin{pmatrix}\mqty{
			0 & 1 & 0 &0 \\
			1 & 0 & 0 &0 \\
			0 & 0 & 0 &0 \\
			0 & 0 & 0 &0 
			}\end{pmatrix},
		\;
		t^2=
			\begin{pmatrix}\mqty{
			0 & -i & 0 &0 \\
			i & 0 & 0 &0 \\
			0 & 0 & 0 &0 \\
			0 & 0 & 0 &0 
			}\end{pmatrix},
		\;
		t^3=
			\begin{pmatrix}\mqty{
			1 & 0 & 0 &0 \\
			0 & -1 & 0 &0 \\
			0 & 0 & 0 &0 \\
			0 & 0 & 0 &0 
			}\end{pmatrix},\\
		t^4&=
			\begin{pmatrix}\mqty{
			0 & 0 & 1 &0 \\
			0 & 0 & 0 &0 \\
			1 & 0 & 0 &0 \\
			0 & 0 & 0 &0 
			}\end{pmatrix},
		\;
		t^5=
			\begin{pmatrix}\mqty{
			0 & 0 & -i &0 \\
			0 & 0 & 0 &0 \\
			i & 0 & 0 &0 \\
			0 & 0 & 0 &0 
			}\end{pmatrix},\\
		t^6&=
			\begin{pmatrix}\mqty{
			0 & 0 & 0 &0 \\
			0 & 0 & 1 &0 \\
			0 & 1 & 0 &0 \\
			0 & 0 & 0 &0 
			}\end{pmatrix},
		\;
		t^7=
			\begin{pmatrix}\mqty{
			0 & 0 & 0 &0 \\
			0 & 0 & -i &0 \\
			0 & i & 0 &0 \\
			0 & 0 & 0 &0 
			}\end{pmatrix},
		\;
		t^8=\frac{1}{\sqrt{3}}
			\begin{pmatrix}\mqty{
			1 & 0 & 0 &0 \\
			0 & 1 & 0 &0 \\
			0 & 0 & -2 &0 \\
			0 & 0 & 0 &0 
			}\end{pmatrix},\\
		t^9&=
			\begin{pmatrix}\mqty{
			0 & 0 & 0 &1 \\
			0 & 0 & 0 &0 \\
			0 & 0 & 0 &0 \\
			1 & 0 & 0 &0 
			}\end{pmatrix},
		t^{10}=
			\begin{pmatrix}\mqty{
			0 & 0 & 0 &-i \\
			0 & 0 & 0 &0 \\
			0 & 0 & 0 &0 \\
			i & 0 & 0 &0 
			}\end{pmatrix},
		t^{11}=
			\begin{pmatrix}\mqty{
			0 & 0 & 0 &0 \\
			0 & 0 & 0 &1 \\
			0 & 0 & 0 &0 \\
			0 & 1 & 0 &0 
			}\end{pmatrix},
		t^{12}=
			\begin{pmatrix}\mqty{
			0 & 0 & 0 &0 \\
			0 & 0 & 0 &-i \\
			0 & 0 & 0 &0 \\
			0 & i & 0 &0 
			}\end{pmatrix},\\
		t^{13}&=
			\begin{pmatrix}\mqty{
			0 & 0 & 0 &0 \\
			0 & 0 & 0 &0 \\
			0 & 0 & 0 &1 \\
			0 & 0 & 1 &0 
			}\end{pmatrix},
		t^{14}=
			\begin{pmatrix}\mqty{
			0 & 0 & 0 &0 \\
			0 & 0 & 0 &0 \\
			0 & 0 & 0 &-i \\
			0 & 0 & i &0 
			}\end{pmatrix},
		t^{15}=\frac{1}{\sqrt{6}}
			\begin{pmatrix}\mqty{
			1 & 0 & 0 &0 \\
			0 & 1 & 0 &0 \\
			0 & 0 & 1 &0 \\
			0 & 0 & 0 &-3 
			} \end{pmatrix}.
	\end{align*}
These generators satisfy an orthonormal condition, $\text{Tr}[T^aT^b]=\delta_{ab}/2$.
	
\newpage
\section{Generalization of one-loop effective potential to fermions in $N$-rank totally symmetric representation}
In this appendix, the calculation of one-loop effective potential is extended to 
the case of $N$-rank totally symmetric representation of $SU(4)$. 
Then, the electroweak symmetry breaking and Higgs mass in this case will be discussed 
whether the results in the $\overline{{\bf 35}}$ representation is improved. 

The one-loop effective potential of fermion in the $N$-rank totally symmetric representation of $SU(4)$ 
can be expressed by using floor and ceiling functions, $\lfloor\quad\rfloor$ and $\lceil\quad\rceil$ as follows. 
	\begin{align}
	V_{\mathrm{eff}}^f=&
	\frac{3}{4\pi^7R^4}\sum_{(k_1,k_2)\neq(0,0)}
	\sum_{n=0}^{\lceil N/2\rceil-1}\sum_{l=0}^{n}
	\left\{l(1+(-1)^{k_2})+1\right\}
	\nu\left(\left(\frac N2-n\right)\alpha_1,\left(\frac N2-n\right)\alpha_2\right)
	\nonumber \\
	&+\frac{3}{4\pi^7R^4}\sum_{(k_1,k_2)\neq(0,0)}
	\sum_{n=0}^{\lfloor N/2\rfloor}\sum_{l=0}^{n}
	l\left(1+(-1)^{k_2}\right)
	\nu\left(\left(\frac{N+1}2-n\right)\alpha_1,\left(\frac{N+1}2-n\right)\alpha_2\right).
	\label{merged}
	\end{align}

We have calculated the above one-loop effective potential and its minimum up to $N=20$ 
and the results up to $N=10$ are explicitly shown in Figs. 4 and 5.   
As we can see, the values of minimum $\alpha_{1,2}$ cannot be of order ${\cal O}(0.01)$ 
required from reproducing a realistic Higgs mass.  
Unfortunately, the results are not so improved.   

	\begin{figure}[H]
		\centering
		\includegraphics[width=150mm]{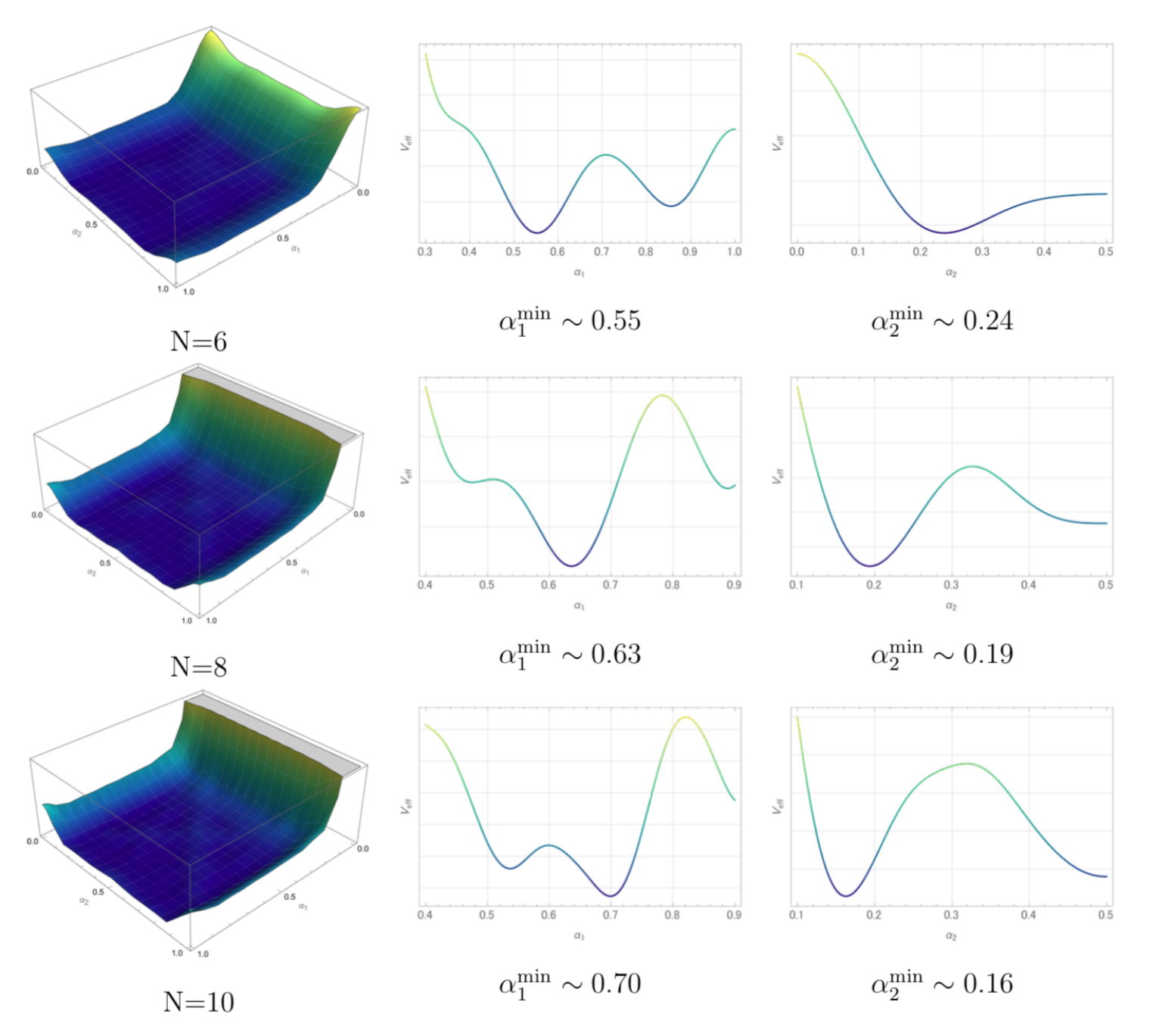}
		\caption{Plots of potential and its minimum for $N = 6, 8, 10$ cases.}
		\label{fig:compactscale:even}
	\end{figure}


	\begin{figure}[H]
		\centering
		\includegraphics[width=150mm]{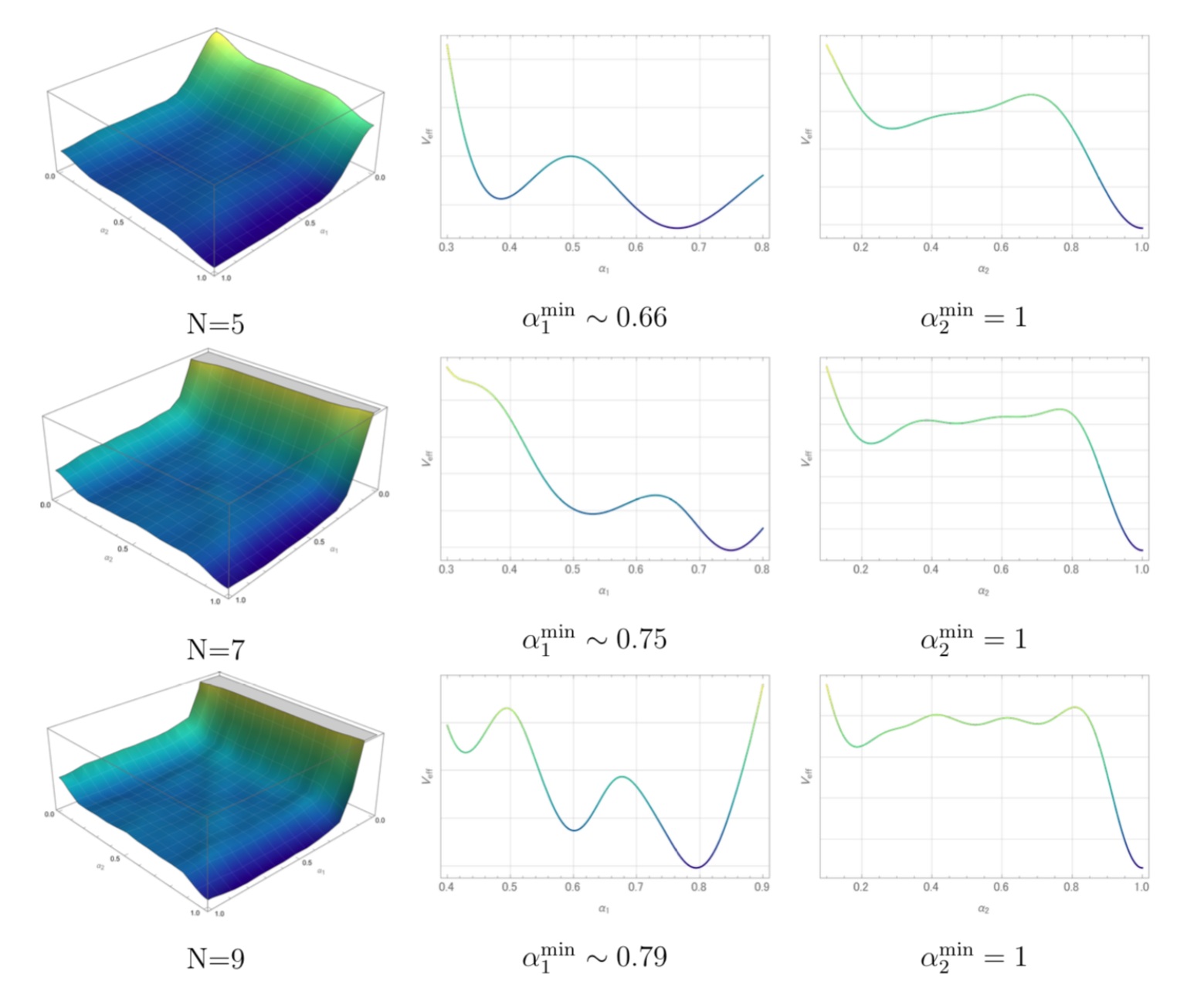}
		\caption{Plots of the potential and minimum for $N = 5, 7, 9$ cases.}
		\label{fig:compactscale:odd}
	\end{figure}




\begin{thebibliography}{104}

\bibitem{2HDMReview}
G.~C.~Branco, P.~M.~Ferreira, L.~Lavoura, M.~N.~Rebelo, M.~Sher and J.~P.~Silva,
``Theory and phenomenology of two-Higgs-doublet models,''
Phys. Rept. \textbf{516}, 1-102 (2012)
[arXiv:1106.0034 [hep-ph]].

\bibitem{2HDMexp}
J.~Haller, A.~Hoecker, R.~Kogler, K.~M\"onig, T.~Peiffer and J.~Stelzer,
Eur. Phys. J. C \textbf{78}, no.8, 675 (2018)
[arXiv:1803.01853 [hep-ph]].


\bibitem{Manton}
N. S. Manton,
``A New Six-Dimensional Approach to the Weinberg-Salam Model,"
Nucl. Phys. B 158 (1979) 141-153.

\bibitem{Fairlie}
  D.~B.~Fairlie,
  ``Higgs' Fields And The Determination Of The Weinberg Angle,''
  Phys.\ Lett.\ B {\bf 82}, 97 (1979); 
  ``Two Consistent Calculations Of The Weinberg Angle,''
  J.\ Phys.\ G {\bf 5}, L55 (1979).

\bibitem{Hosotani1}
Y.Hosotani, 
``Dynamical Mass Generation by Compact Extra Dimension,"
Phys. Lett. \textbf{126B} (1983) 309.

\bibitem{Hosotani2}
Y.Hostonani,
``Dynamical of Non-integrable Phases and Gauge Symmetry Breaking,"
ANNALS \textbf{190}, (1989) 233-253


\bibitem{HIL}
H.~Hatanaka, T.~Inami and C.S.~Lim,
  ``The Gauge hierarchy problem and higher dimensional gauge theories,''
  Mod.\ Phys.\ Lett.\ A {\bf 13}, 2601 (1998);
  [hep-th/9805067].
  
\bibitem{ABQ}
I.~Antoniadis, K.~Benakli and M.~Quiros,
``Finite Higgs mass without supersymmetry,''
  New J.\ Phys.\  {\bf 3}, 20 (2001);
  [arXiv:hep-th/0108005].
  

\bibitem{MY}
N.~Maru and T.~Yamashita,
 ``Two-loop calculation of Higgs mass in gauge-Higgs unification:
  5D  massless QED compactified on S**1,''
  Nucl.\ Phys.\ B {\bf 754}, 127 (2006);
  [arXiv:hep-ph/0603237].

\bibitem{HMTY}
  Y.~Hosotani, N.~Maru, K.~Takenaga and T.~Yamashita,
 ``Two loop finiteness of Higgs mass and potential in the gauge-Higgs unification,''
  Prog.\ Theor.\ Phys.\  {\bf 118}, 1053 (2007);
  [arXiv:0709.2844 [hep-ph]].

 \bibitem{LMH}
  C.S.~Lim, N.~Maru and K.~Hasegawa,
 ``Six dimensional gauge-Higgs unification with an extra space S**2 
 and the hierarchy problem,''
    J.\ Phys.\ Soc.\ Jap.\  {\bf 77}, 074101 (2008);
  arXiv:hep-th/0605180. 
  
\bibitem{HLM}
K.~Hasegawa, C.~S.~Lim and N.~Maru,
``Predictions of the Higgs mass and the weak mixing angle in the 6D gauge-Higgs unification,''
J. Phys. Soc. Jap. \textbf{85}, no.7, 074101 (2016); 
[arXiv:1509.04818 [hep-ph]].


\bibitem{SSSW}
C.~A.~Scrucca, M.~Serone, L.~Silvestrini and A.~Wulzer,
``Gauge Higgs unification in orbifold models,''
JHEP \textbf{02}, 049 (2004); 
[arXiv:hep-th/0312267 [hep-th]].

\bibitem{CGM}
C.~Csaki, C.~Grojean and H.~Murayama,
``Standard model Higgs from higher dimensional gauge fields,''
Phys. Rev. D \textbf{67}, 085012 (2003);  
[arXiv:hep-ph/0210133 [hep-ph]].

\bibitem{MS}
Y.~Matsumoto and Y.~Sakamura,
``6D gauge-Higgs unification on T$^{2}$ /Z$_{N}$ with custodial symmetry,''
JHEP \textbf{08}, 175 (2014)
[arXiv:1407.0133 [hep-ph]].

\bibitem{HNT}
Y.~Hosotani, S.~Noda and K.~Takenaga,
``Dynamical gauge-Higgs unification in the electroweak theory,''
Phys. Lett. B \textbf{607}, 276-285 (2005)
[arXiv:hep-ph/0410193 [hep-ph]].

\bibitem{SSS}
C.~A.~Scrucca, M.~Serone and L.~Silvestrini,
``Electroweak symmetry breaking and fermion masses from extra dimensions,''
Nucl. Phys. B \textbf{669}, 128-158 (2003); [arXiv:hep-ph/0304220 [hep-ph]].

\bibitem{CCP}
G.~Cacciapaglia, C.~Csaki and S.~C.~Park,
  ``Fully radiative electroweak symmetry breaking,''
  JHEP {\bf 0603}, 099 (2006); 
  [arXiv:hep-ph/0510366]. 
  
\end{thebibliography}
\end{document}